\documentclass[amssymb, nofootinbib,superscriptaddress]{revtex4}

\usepackage{amsmath}
\usepackage{amsfonts}
\usepackage{graphicx}
\usepackage{psfrag}

\newcommand{\N}{\mathbb{N}}

\newcommand{\cA}{{\mathcal A}}
\newcommand{\cE}{{\mathcal E}}
\newcommand{\cF}{{\mathcal F}}
\newcommand{\cM}{{\mathcal M}}
\newcommand{\cS}{{\mathcal S}}
\newcommand{\cT}{{\mathcal T}}
\newcommand{\cZ}{{\mathcal Z}}
\renewcommand{\t}{\tilde}

\newcommand{\Sdim}{\mathrm{Sdim}}

\newcommand{\R}{\mathbb{R}}
\newcommand{\C}{\mathbb{C}}
\renewcommand{\N}{\mathbb{N}}

\renewcommand{\u}{\mathfrak{u}}
\newcommand{\su}{\mathfrak{su}}
\newcommand{\osp}{\mathfrak{osp}}
\renewcommand{\o}{\mathfrak{o}}
\renewcommand{\sp}{\mathfrak{sp}}

\newcommand{\gde}{\delta}

\newcommand{\gla}{\lambda}

\newcommand{\man}{\cM}
\newcommand{\dman}{\Delta}

\renewcommand{\dim}{\mathrm{dim}}

\newcommand{\tr}{\mathrm{tr}}
\newcommand{\Str}{\mathrm{Str}}
\newcommand{\w}{\wedge}
\newcommand{\pd}{\partial}

\newcommand{\be}{\begin{equation}}
\newcommand{\ee}{\end{equation}}
\newcommand{\bes}{\begin{eqnarray}}
\newcommand{\ees}{\end{eqnarray}}
\begin{document}
%

\title{\Large\bf $N=2$ supersymmetric spin foams in three dimensions}


\author{Etera R. Livine}\email{etera.livine@ens-lyon.fr}
\affiliation{Laboratoire de Physique, ENS Lyon, CNRS UMR 5672, 46 All\'ee d'Italie, 69007 Lyon, France}

\author{James P. Ryan}\email{jryan@perimeterinstitute.ca}
\affiliation{Perimeter Institute, 31 Caroline St. N., Waterloo, ON N2L 2Y5, Canada}

%
%
%

\date{\small October 18, 2007}
%
%

\begin{abstract}
We construct the spin foam model for $N = 2$ supergravity in three dimensions.  Classically, it is
a $BF$ theory with gauge algebra $\osp(2|2)$.  This algebra has representations which are not
completely reducible.  This complicates the procedure when building a state sum.  Fortunately, one
can and should excise these representations. We show that the restricted subset of representations
form a subcategory closed under tensor product.  The resulting state-sum is once again a
topological invariant.   Furthermore, within this framework one can identify positively and
negatively charged fermions propagating on the spin foam. These results on $\osp(2|2)$
representations and intertwiners apply more generally to spin network states for  $N=2$ loop
quantum supergravity (in $3+1$ dimensions) where it allows to define a notion of BPS states.
\end{abstract}

\maketitle

%
\section{Introduction}

Supergravity and supersymmetry have enjoyed widespread popularity and attracted intensive study in
the general high energy physics community.   They have, however,  been relatively under-appreciated
in the arena of non-perturbative quantum gravity, with the exception of \cite{Ling:1999gn,
Ling:2000ss, Ling:2000su}  on canonical side and \cite{Livine:2003hn} for spin foams in 3
dimensions.

In \cite{Livine:2003hn}, they provide a general framework for the quantization of supersymmetric
theories, with emphasis placed on $N=1$ super-$BF$ theory with gauge algebra $\osp(1|2)$.  This
arises as a particular example of a super Chern-Simons theory, proposed classically in
\cite{Achucarro:1987vz},  in the limit where the cosmological constant goes to zero.   The
quantization technique is based upon the $\su(2)$ Ponzano-Regge model for quantum gravity in three
dimensions.    It is a spin foam model regularizing in a cut-off independent way the $BF$-path
integral.  Furthermore, representations of $\osp_E(1|2)$ \footnote{We differentiate the cases of
Riemannian and Lorentzian supergravity theories with the subscripts $E$ and $L$, respectively.} are
formed from the direct sum of two irreducible $\su(2)$ representations.  Therefore, the resulting
$\osp_E(1|2)$ model contains the $\su(2)$ Ponzano-Regge state sum nested within.  The other
configurations, referred to as its superpartners, can be identified with fermions propagating along
the edges of the spin foam.  A plausible argument was given to this interpretation of fermions
propagating and interacting on a dynamical background,  but it was not developed explicitly.  Only
an asymptotic analysis of the spin foam amplitudes, and a subsequent comparison with configurations
of super-Regge calculus would provide the necessary reinforcement.  We shall review this in Section
\ref{susyone}.

Our quest here is not to investigate issues pertaining to the semi-classical regime.  We leave that
for later work.  But we want to extend the formalism to the $N=2$  scenario.   Extended
supersymmetry is essential to the success of string theory, which investigates quantum gravity in
the perturbative regime.  A consistent spin foam theory with $N=2$ supersymmetry allows one to
examine the properties of BPS states in a non-perturbative setting, and it could provide a way to
compare results on the black hole entropy got by both string theory and loop quantum gravity.  The
difficulty facing us when trying to quantize $\osp(2|2)$ $BF$ theory is that the representation
theory of the algebra is highly non-trivial. The representations are not all distinguishable using
the two Casimirs.  Moreover, some of these representations are not even completely reducible.
Propitiously, excising these representations (among others), we arrive at a subcategory which is
closed under tensor product, and which is exactly the subset of representations upon which one can
define a star product \cite{Scheunert:1976wj}. Then such a category of representations stable under
tensor product allows to define a topological state sum model \cite{topoinv}.
%
As we will see,  pure $N=2$ supergravity naturally incorporates two charged fermions  and a
(topological) $\u(1)$ gauge field.  Thus, the spin foam model will contain configurations
interpretable as positively and negatively charged fermions propagating along its edges.  Similar
matters have recently come under investigation \cite{Freidel:2004vi, Freidel:2004nb,
Freidel:2005bb, Fairbairn:2006dn, Speziale:2007mt, Oriti:2002bn}, and have received much attention in the case of
spin foam quantum gravity.

\section{Review of $N=1$ supersymmetric spin foams}
\label{susyone}

For a thorough investigation of $N=1$ supersymmetric spin foams we refer the reader to
\cite{Livine:2003hn}.  We start from a $BF$-type action for $3d$ supergravity with zero
cosmological constant
\be
\cS[\cE,\cA]=\int_{\man} \Str(\cE\w\cF[\cA])
\ee
where $\cE$ is the supertriad, $\cA$ is the superconnection, while $\cF[\cA] = d\cA + \cA \w \cA$
is the supercurvature.  Both $\cE$ and $\cA$ are 1-forms valued in the super lie algebra, and
$\Str$ is its supertrace. In the case of Riemannian supergravity, this gauge algebra is
$\osp_E(1|2)$.  Its bosonic subalgebra is $\su(2)$. It is a minimal supersymmetric extension of
$3d$ gravity.

As a brief aside, we may easily add a positive cosmological constant term to the action and in this
context the action is equivalent to a super Chern-Simons theory devised by Ach\'ucarro and Townsend
\cite{Achucarro:1987vz}. The theory in question is Riemannian deSitter supergravity with gauge
algebra $\osp_E(1|2)\times\osp_E(1|2)$.

Apart from its $\su(2)$ subalgebra, $\osp_E(1|2)$ has fermionic (anti-commuting) generators
$Q_{\pm}$.   Together with $J_\pm,\;J_3$, they satisfy the algebra
\be
\begin{array}{lcll}
[J_3,J_\pm] = \pm J_\pm, &\quad  & [J_+,J_-] = 2J_3,& \\

[J_3,Q_\pm]= \pm\frac{1}{2} Q_\pm, &\quad &   [J_\pm,Q_\pm] = 0,  &  [J_\pm, Q_\mp] = Q_\pm,\\

 \{Q_\pm,Q_\pm\} = \pm \frac{1}{2} J_\pm,&\quad &  \{Q_\pm, Q_\mp\} = -\frac{1}{2} J_3&
\end{array}
%
%
\ee
The supergravity fields written in terms of generators of the algebra are
\be
\cE = E^iJ_i + \phi^A Q_A, \qquad \cA = W^iJ_i + \psi^A Q_A,\\
\ee
where $E$ and $W$ are the triad and connection, while $\phi$ and $\psi$ represent the fermion
field.   $A\in\{\pm\}$ and $i\in\{1,2,3\}$.  The action may be rewritten in terms of these
variables as
\be
\cS_{N = 1}[E,W,\phi,\psi] = \int_{\man} \Big\{\Str\big(E\w (F[W] + \psi\wedge\psi)\big) + \phi \w D\psi)\Big\}
\ee
where $F(W) = dW + W\w W$ is the gravitational curvature, and we define the operator as $D = d +
W\w$.  This action describes a fermion field propagating on a manifold $\man$ endowed with a
dynamical geometry.\footnote{The spinor indices follow the north-west convention so that $\phi^{A}
= \epsilon^{AB}\phi_B$ and $\phi_{A} = \phi^B\epsilon_{BA}$.  The metric on the spinor space is the
anti-symmetric tensor  $\epsilon_{AB}$ with $\epsilon_{+-} = \epsilon^{+-} = 1$.  This implies
$\epsilon^{AB}\epsilon_{BC} = - \delta^A_{\;B}$.
The quadratic Casimir of $\osp_E(1|2)$ determines its supertrace.  This takes the form $C_2=
J_{i\;}\eta^{ij}J_j + Q_{A\;} \epsilon^{AB} Q_B$, where $\eta_{ij} = \delta_{ij}$.  Thus
$\Str(J_iJ_j) = \eta_{ij}$, $\Str(Q_AQ_B) = -\epsilon_{AB}$, and $\Str(J_iQ_A) = 0$.   }

The recipe for quantization bases itself on the Ponzano-Regge (PR) model for $3d$ Riemannian
quantum gravity \cite{Ponzano:1968}.  The PR model is a discrete state sum, derivable from the
$\su(2)$-$BF$ path integral.   The method is to triangulate the manifold $\cM$ using a simplicial
complex $\dman$.    The gravitational information is encoded in the representations of $\su(2)$
which label the edges $e$ of $\dman$, denoted $V^{j_e}$.  The amplitude assigned to the triangular
faces $f$ of each tetrahedron in $\dman$ are invariant tensors.  They are the $\su(2)$ $3j$-symbols
intertwining the three edge representations of the triangle, denoted $i_f:V^{j_1}\otimes
V^{j_2}\otimes V^{j_3}\rightarrow \mathbb{C}$ or $i_f:V^{j_1}\otimes V^{j_2}\otimes
(V^{j_3})^*\rightarrow \mathbb{C}$ depending on the relative orientation of the edges.\footnote{The
state sum includes both orientations for each edge.}  The partition function arises by applying
this procedure to  a closed manifold $\cM$, and takes the form

\begin{figure}[h]
\centering
\hspace{1.5cm}\begin{minipage}[r]{0.54\linewidth}
$$\cZ_{BF} [\dman] = \sum_{V^{j_e}} \prod_{e} \dim(V^{j_e}) \prod _{t} \cT(V^{j_e}),\quad\textrm{where}\quad \cT(V^{j_e}) = $$
\end{minipage}
\begin{minipage}[l]{0.2\linewidth}
\psfrag{A}{\tiny $f_1$}
\psfrag{B}{\tiny $f_2$}
\psfrag{C}{\tiny $f_3$}
\psfrag{D}{\tiny $f_4$}
\psfrag{a}{\tiny$e_1$}
\psfrag{b}{\tiny $e_2$}
\psfrag{c}{\tiny $e_3$}
\psfrag{d}{\tiny $e_4$}
\psfrag{e}{\tiny $e_5$}
\psfrag{f}{\tiny $e_6$}
\includegraphics[width = \linewidth]{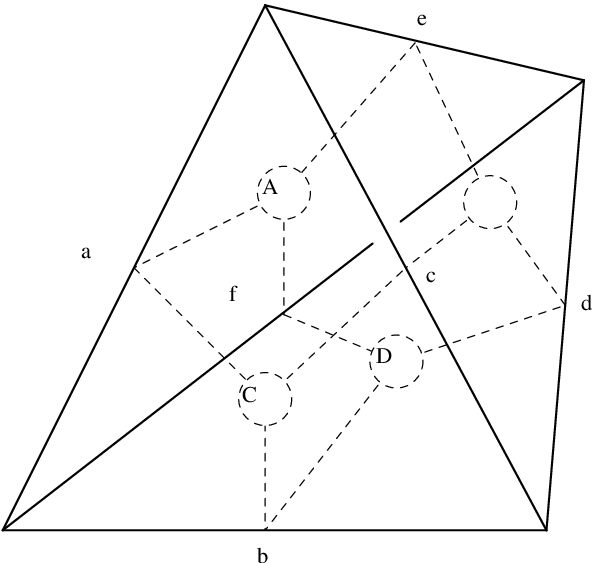}
\end{minipage}\hspace{1.5cm}
\begin{minipage}[r]{0.05\linewidth}
\be\label{part1}\ee
\end{minipage}
\end{figure}

\noindent
where $e$ and $t$ are the edges and tetrahedra of $\dman$ respectively.   $V^{j_e}$ is the
irreducible representation of $\su(2)$ labeled by $j_e\in\frac{1}{2}\N$.  $\dim(V^{j_e}) = 2j_e+1$
is the dimension of $V^{j_e}$.  Finally, $\cT(V^{j_e}) = \{6j\}_{\su(2)}$ is the $\{6j\}$-symbol
for $\su(2)$. It is the amplitude for the tetrahedron and it arises from a contraction of the
intertwiners assigned to the four faces which bound the tetrahedron $t$. In (\ref{part1}), we also
drew the contracted intertwiners as a trivalent graph on the boundary of the tetrahedron.  This is
known as the boundary spin network.   Another important structure is the dual 2-skeleton of the
triangulation denoted $\dman_2^*$.  We can equivalently think of the representations as labeling
the faces $f^*$ of $\dman_2^*$ and the intertwiners as labeling the edges $e^*$ of $\dman_2^*$.
This is the structure known as a spin foam.

The resulting amplitude is a topological invariant, and thus is independent of the particular
triangulation one chooses initially for $\cM$.\footnote{Often we are interested in transitions
amplitudes between two quantum states. For this we need a manifold with boundary $\cS$. A quantum
state is a spin network representing the gravitational information residing on the boundary $\cS$.
A spin network is a graph labeled with the boundary gravitational information, and is dual to a
triangulation $\dman_{\cS}$  of $\cS$.    Then, $\dman$ should coincide with $\dman_{\cS}$ on the
boundary.  The resulting amplitude is a topological invariant up to the boundary contributions.}
This is as it should be, since $3d$ gravity lacks local degrees of freedom propagating in the bulk.

One hopes na\"ively,  that by swapping $\su(2)$ for $\osp_E(1|2)$ one can arrive at a topological
state-sum for $N = 1$ quantum supergravity.  The representations of $\osp_E(1|2)$ can be decomposed
over its bosonic subalgebra.  Each representation $R^{j}$ of $\osp_E(1|2)$ comprises of the direct
sum of two representations of $\su(2)$.\footnote{The action of the operators on  the representation
$R^{j}$ is
\be
\begin{array}{rlrl}
J_3|j,j,m> \;=& m|j,j,m>,  & J_3|j,j-\frac{1}{2}, m>\; = & m|j,j,m>,\\
J_\pm |j,j,m>\; =& \sqrt{(j\mp m)(j\pm m + 1)} |j,j, m\pm 1>, & J_\pm |j,j-\frac{1}{2},m>\; =& \sqrt{(j-\frac{1}{2}\mp m)(j+\frac{1}{2}\pm m)} |j,j-\frac{1}{2}, m\pm 1>,  \\
Q_{\pm} |j,j,m>\;  = & \mp\sqrt{j\mp m}|j,j-\frac{1}{2}, m\pm \frac{1}{2}>,  & Q_{\pm} |j, j-\frac{1}{2}, m> \; = & -\frac{1}{2}\sqrt{j+\frac{1}{2} \pm m}\;|j,j,m\pm \frac{1}{2}>
\end{array}
\ee
while the Casimir of the representation is $C^j = j\big(j+\frac{1}{2}\big)$.} Thus the
representations of $\osp_E(1|2)$ are once again labeled by $j\in\frac{1}{2}\N$ with
$j\geq\frac{1}{2}$.  Furthermore, the tensor product of two representations of $\osp_E(1|2)$
satisfy a rule analogous to that of $\su(2)$ except that the sum over $j$ goes in {\it
half-integer} steps
\be
R^{j} = V^j \oplus V^{j-\frac{1}{2}}, \phantom{xxxxxxxx} R^{j_1}\otimes R^{j_2} = \bigoplus_{|j_1 - j_2| \leq j \leq j_1 + j_2} R^{j}.
\ee

In the case of $\osp_E(1|2)$, we cannot simply define a star operator and talk about unitary
representations.  To construct unitary representations, we need a grade star operator.  We obtain
this by labeling each irreducible representation $R^j$ with a parity $\gla = 0, 1$, so that the
representation is decomposed as $R^{j,\gla} = V^{j,\gla}\oplus V^{j-\frac{1}{2}, \gla+1}$.  As it
stands, we have two copies of each representation, one for each choice of parity, but if the
subalgebra $\su(2)$ is to play the physical role of rotations, the representations must obey the
spin-statistics relation of quantum field theory. In other words, the $V^j$ should be even or odd
depending on whether $j$ is an integer or not.  The implications are
\begin{displaymath}
\begin{array}{llll}
\textrm{For $j\in \N$}, \quad & Q^{\dagger}_+ = -Q_-, \quad &\mathrm{and} \quad & Q^{\dagger}_- = Q_+,\\
\textrm{for $j\in \N+\frac{1}{2}$}, \quad & Q^{\dagger}_+ = Q_-, \quad &\mathrm{and} \quad & Q^{\dagger}_- = -Q_+.\\
\end{array}
\end{displaymath}
It is this parity issue that complicates the simple replacement of $\su(2)$ by $\osp_E(1|2)$.
Unlike the $\su(2)$ case, the $3j$-symbols of $\osp_E(1|2)$ no longer have a purely combinatorial
definition, but also depend on their graphical representation.  This dependence is familiar from
quantum groups.   The appropriate dependence can be incorporated into a graphical calculus known as
circuit diagrams described in \cite{topoinv,Livine:2003hn}.\footnote{A circuit diagram consists of
lines called wires, and boxes called cables. The circuit diagram is an enrichment of the dual
2-skeleton $\dman_2^*$.  Each edge $e^*$ of $\dman_2^*$ is replaced by a cable through which three
wires pass. Each wire inherits the representation of one of the three incident faces $f^*$ of
$\dman^*$.  The 4-valent vertices of $\dman_2^*$ are replaced by a routing of the twelve incident
wires, such that the faces $f^*$ of $\dman_2^*$ are replaced by a closed loop.  The non-trivial
rule which can now be encoded is that should two wires cross in the planar embedding of the circuit
diagram, then one should include a factor of $(-1)^{\lambda_j \lambda_k}$ where the $\lambda$ are
the parities of the involved representations.
}  Taking into account this dependence, we arrive at the state sum model
\be
\cZ_{N=1}[\dman] = \sum_{R^{j_e}} \prod_{e} \Sdim(R^{j_e}) \prod _{t} \cT(R^{j_e}),
\ee
where $\cT(R^{j_e}) = \{6j\}_{\osp_E(1|2)}$ is the $\{6j\}$-symbol for  $\osp_E(1|2)$.  The
superdimension of the representation is given by the supertrace of the identity element in that
representation
\be
\Sdim(R^j) = (-1)^{2j}\dim(V^j) - (-1)^{2j}\dim(V^{j-\frac{1}{2}}) = (-1)^{2j}.
\ee
A representation of $\osp_E(1|2)$ labels each edge. The four triangles of each tetrahedron in
$\dman$ are each labeled by a trivalent intertwiner  $I_f$ of $\osp_E(1|2)$. A contraction of these
intertwiners gives rise to $\{6j\}_{\osp_E(1|2)}$.
This defines a topological state sum as proved in \cite{Livine:2003hn} following the circuit
diagram techniques introduced in \cite{topoinv}. It is interpreted as providing the spinfoam
quantization of $N=1$ 3d supergravity.

We shall be more precise.  To construct the amplitude, we first label each edge of $\dman$ with an
orientation. This allows us to distinguish a representation from its dual.  (In a PR-like model,
one sums over both orientations.)   Then we consider a tetrahedron $t$ in $\dman$.  We assign an
intertwiner to each of its faces: $I_f: R^{j_1}\otimes R^{j_2}\otimes R^{j_3}\rightarrow \C$ or
$I_f: R^{j_1}\otimes R^{j_2}\otimes (R^{j_3})^*\rightarrow \C$ depending on the relative
orientation of the three edges of that face.  This intertwiner is unique up to normalisation and we
may normalise it by imposing that the complete contraction of two intertwiners equals the identity.
Contracting these four intertwiners produces the $\{6j\}_{\osp_E(1|2)}$ amplitude for the
tetrahedron.

\begin{figure}[h]
\begin{center}
\psfrag{A}{$R^{j_1}$}
\psfrag{B}{$R^{j_2}$}
\psfrag{C}{$R^{j_3}$}
\psfrag{=}{$=$}
\psfrag{narcotango}{$j_1+j_2+j_3\in\mathbb{N}$}
\psfrag{nuevotango}{$j_1+j_2+j_3\in\mathbb{N}+\frac{1}{2}$}
\includegraphics[width=10cm]{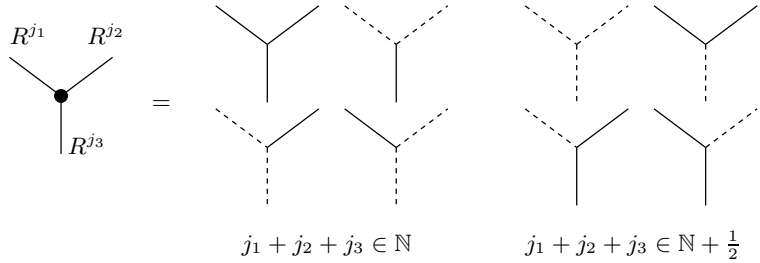}
\caption{\label{1threej} The $\osp_E(1|2)$ intertwiner. The isospin spaces $V^{j_i}$ are denoted by solid lines and $V^{j_i-\frac{1}{2}}$ by dotted lines.}
\end{center}
\end{figure}

Applying the isospin decomposition to the intertwiner of $\osp_E(1|2)$ allows us to distinguish
between two cases, as shown in Figure \ref{1threej}.
Four of these intertwiners satisfy $j_1+j_2+j_3 \in \N$.  These occur in the $\su(2)$ Ponzano-Regge
state sum.  The other four satisfy $j_1+ j_2+j_3\in\N+\frac{1}{2}$.  These $\su(2)$ intertwiners
inherit their normalisation from that of the $\osp_E(1|2)$ intertwiner.    The interpretation
proposed in \cite{Livine:2003hn} for these intertwiners is that the first four are bosonic and
represent pure gravity, while the second four (which do not occur in the $\su(2)$ Ponzano-Regge
model) are fermionic intertwiners and denote the presence of a fermion.  Importantly, the two
$\osp_{E}(1|2)$ intertwiners attached to the shared face of two adjacent tetrahedra are the same.
Thus, they are both bosonic or both fermionic (although they need not be exactly the same bosonic
or fermionic). Therefore, the fermions may be thought of as propagating along the edges of the dual
2-skeleton $\dman_2^*$.

\begin{figure}[h]
\psfrag{A}{\tiny $F$}
\psfrag{B}{\tiny $G$}
\psfrag{C}{\tiny $F$}
\psfrag{D}{\tiny $G$}
\psfrag{=}{\footnotesize $=$}
\psfrag{+}{\footnotesize $+$}
\psfrag{do}{\footnotesize$\cdots$}
\includegraphics[width = 16cm]{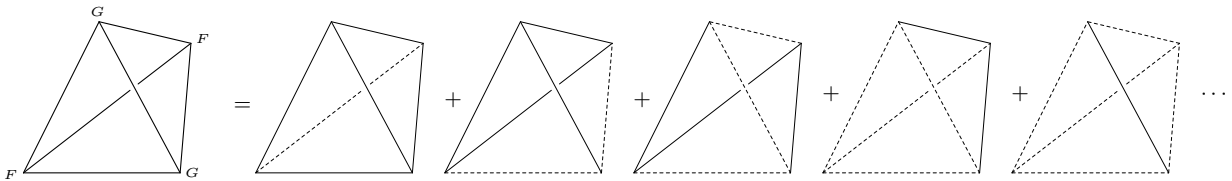}
\caption{\label{sixj}
The $(G,G,F,F)$ class of diagrams.  $G$  stands for bosonic intertwiners, while $F$ represents
fermionic intertwiners.  There are five diagrams occurring in this decomposition up to
permutations.}
\end{figure}

Again, due to the isospin decomposition of the representations, we can distinguish between
different types of tetrahedral amplitude, which are formed from a contraction of the intertwiners
$I_f$ assigned to the four triangles of a tetrahedron. There are many possible terms, but they fall
into three classes:  those consisting of four bosonic intertwiners $(G,G,G,G)$, two fermionic and
two bosonic intertwiners $(G,G,F,F)$, and four fermionic intertwiners $(F,F,F,F)$.   The cases with
an odd number of fermionic intertwiners have zero amplitude.   For example,  we illustrate the
class of diagrams (up to permutation) occurring in the $(G,G,F,F)$  tetrahedron in Figure
\ref{sixj}.  The tetrahedra drawn in the diagram are not a space-time tetrahedra but each are the
contraction of four intertwiners (the dual spin network).   Note that dotted lines either join the
fermion or form a closed loop.  This is a generic feature of any boundary spin network and was
proven in \cite{Livine:2003hn}.

\noindent
Through this construction we have identified fermionic degrees of freedom attached to the
intertwiners,  and interpreted the $\osp_E(1|2)$ Ponzano-Regge model as providing a path integral
for gravity plus fermions.  The fermionic edges of the dual 2-skeleton form a Feynman graph on the
dynamical spin foam.  The interpretation of the amplitudes given here is internally consistent but
needs to be solidified by analysing, for example, the semiclassical regime in which one could
relate the amplitudes to those coming from Regge calculus coupled to spin-$\frac{1}{2}$ fermions
\cite{Livine}.

\section{N=2 supergravity and supersymmetric spin foams}
\label{susytwo}

The analysis of Ach\'ucarro and Townsend \cite{Achucarro:1987vz} extends the Chern-Simons formalism
for three dimensional gravity to a whole class of supersymmetric cases.  The symmetry algebra of
Lorentzian Anti de Sitter gravity is $\sp(2)\oplus\sp(2)$.  Replacing this by its supersymmetric
counterpart $\osp_L(p|2)\oplus\osp_L(q|2)$, we arrive at  theory known as $(p,q)$ AdS supergravity.
An analogous avenue can be followed for Euclidean de Sitter gravity where we replace
$\su(2)\oplus\su(2)$ by $\osp_E(p|2)\oplus\osp_E(q|2)$.\footnote{Note that the bosonic sector of
$\osp_L(p|2)$ is $\o(p)\oplus\sp(2)$ while the bosonic sector of $\osp_E(p|2)$ is
$\o(p)\oplus\su(2)$.}

The theory in which we shall be interested is $(2,2)$ Riemannian dS supergravity, that is $p = 2$
and $q = 2$.  Similar to case of $N=1$ Chern-Simons supergravity, we can write this theory as a
$BF$-theory with cosmological constant term based on the gauge algebra $\osp_E(2|2) \oplus
\osp_E(2|2) $.  Taking the limit where the cosmological constant vanishes one arrives at the action
\be
\cS[\cE,\cA]=\int_{\cM} \Str(\cE\w\cF(\cA)).
\ee
The $N=2$ supertriad and superconnection fields are defined as
\be
\cE  =  E^iJ_i + \phi^AQ_A + \t{\phi}^A\t{Q}_A + eB, \hspace{1cm}   \cA  =  W^iJ_i + \psi^AQ_A + \t{\psi}^A\t{Q}_A + wB,
\ee
where $E$ and $W$ are the $\su(2)$-valued triad and connection respectively. $\phi$ and $\psi$
contain the degrees of freedom of the positively charged fermion, while $\t{\phi}$ and $\t{\psi}$
represent the negatively charged fermion. $e$ is the electric field for the $\u(1)$ gauge theory
while $w$ is the $\u(1)$ connection. Finally, $\cF[\cA] = d\cA +\cA\wedge\cA$. The Lie algebra
elements $J_i, Q_\pm, \t{Q}_\pm, B$ satisfy the commutation relations:
\be
\label{susy2cc}
\begin{array}{llll}
[J_+,J_-]=2J_3, & [J_3,J_{\pm}]={\pm}J_{\pm}, & [J_{\pm},B]=0, & [J_3,B]=0,\\
& & &\\

[B,Q_{\pm}]=\frac{1}{2}Q_{\pm}, & [B,\tilde{Q}_{\pm}]=-\frac{1}{2}\tilde{Q}_{\pm}, &[J_3,Q_{\pm}]=\pm\frac{1}{2}Q_{\pm},&[J_3,\tilde{Q}_{\pm}]=\pm\frac{1}{2}\tilde{Q}_{\pm},\\
&&&\\

[J_{\pm},Q_{\mp}]=Q_{\pm},&[J_{\pm},\tilde{Q}_{\mp}]=\tilde{Q}_{\pm},&[J_{\pm},Q_{\pm}]=0,&[J_{\pm},\tilde{Q}_{\pm}]=0,\\
&&&\\

\{Q_{\pm},Q_{\pm}\}=0,&\{Q_{\pm},Q_{\mp}\}=0,&\{\tilde{Q}_{\pm},\tilde{Q}_{\pm}\}=0,&\{\tilde{Q}_{\pm},\tilde{Q}_{\mp}\}=0,\\
&&&\\

&\{Q_{\pm},\tilde{Q}_{\pm}\}={\pm}J_{\pm},&\{Q_{\pm},\tilde{Q}_{\mp}\}=-J_3{\pm}B.&\\
\end{array}
\ee
The generators are of two types, bosonic and fermionic \cite{Scheunert:1976wj}. The bosonic sector
is generated by: $J_{\pm}$, $J_3$, $B$; it is an $\su(2)\oplus\u(1)$ subalgebra.  $\su(2)$ is often
called isospin and $\u(1)$ is the charge. $Q_\pm$ and $\t{Q}_\pm$ are fermionic
generators.\footnote{The map
\be
J_i\rightarrow J_i, \quad\quad B\rightarrow -B, \quad\quad Q_{\pm}\rightarrow\tilde{Q}_{\pm}, \quad\quad \tilde{Q}_{\pm}\rightarrow{Q_{\pm}}
\ee
is an automorphism of the algebra. The quadratic Casimir for the algebra is given by
\be
C_2=J^2-B^2+\frac{1}{2}(Q_+\tilde{Q}_- -Q_-\tilde{Q}_+ -\tilde{Q}_+Q_- -\tilde{Q}_-Q_+).
\ee
Thus, the inner product on the algebra is given by
\be
\tr(J_iJ_j)=\gde_{ij}, \quad\quad \tr(BB)=-1, \quad\quad
\tr(Q_+\tilde{Q}_-)=-2, \quad\quad \tr(\tilde{Q}_+Q_-)=-2
\ee
There also exists a cubic Casimir $C_3$.}

Written out in components of the multiplets $\cE$ and $\cA$, the action can be rewritten as
\be
\cS_{N=2}[\cE, \cA] =
\int_{\cM} \left\{ \Str(E\w (F[W] + \psi\wedge\t{\psi})
+ e\w (f[w] + \psi\wedge\t{\psi}) + \t{\phi}\w D\psi +\phi \w D\t{\psi} \right\}
\ee
where $D = \pd  + W + w$ is the covariant derivative with respect to gravity and the $\u(1)$ gauge
theory. $F(W) = dW + W\w W$ is the gravitational curvature and $f(w) = dw$ is the $\u(1)$ gauge
curvature. We note here that the gauge theory is a $\u(1)$ $BF$-theory rather than electrodynamics
in three dimensions.  Through this mechanism, the fermions acquire a charge but we do not see
chargeless spin-1 particles propagating and interacting in our theory.

The state sum model will be a topological theory based on the representations $R^{j,b}$ of the
superalgebra $\osp_E(2|2)$. In the framework of spin foam models, it should implement a path
integral for gravity plus charged fermions in three dimensions.     We follow the same procedure as
in Section \ref{susyone}. The resulting amplitude is
\be
\label{part2}
\cZ_{N=2} = \sum_{R^{j_e,b_e}} \; \prod_e\Sdim(R^{j_e,b_e})\; \prod_v \cT(R^{j_e,b_e})
\ee
There are several differences between the case of $\osp_E(2|2)$ and $\osp_E(1|2)$.  We wish to
assign a representation to each edge of $\dman$, but to add a complication, the representation
theory is plagued by representations which are not completely reducible.  More explicitly, the
representations of $\osp_E(2|2)$ may written as the sum of representations of its bosonic
subalgebra $\su(2)\oplus\u(1)$.   The representations $V^{j,b}$ of $\su(2)\oplus\u(1)$ are labeled
by an isospin $j\in\frac{1}{2}\N$ and a charge $b\in\C$.

The representations fall into four categories
\begin{itemize}
\item Typical irreducible - $j\neq\pm b$: The typical representations consist of four multiplets of the bosonic subalgebra $\su(2)\oplus\u(1)$
$$R^{j,b}=V^{j,b}\oplus V^{j-\frac{1}{2},b-\frac{1}{2}}\oplus V^{j-\frac{1}{2},b+\frac{1}{2}}\oplus V^{j-1,b}.$$
where $V^{j,b}$ is the representation of $\su(2)\oplus\u(1)$ with quantum numbers $j$ and $b$
referring to $\su(2)$ and $\u(1)$, respectively.  The dimension of such a representation is $8j$.
\item Atypical irreducible - $j=b$: The atypical representations consist of two multiplets of $\su(2)\oplus\u(1)$
$$R^{j,j}=V^{j,j}\oplus V^{j-\frac{1}{2},j+\frac{1}{2}}.$$
The dimension of the representation is $4j+1$.
\item Atypical irreducible - $j=-b$:
$$R^{j,j}=V^{j,-j}\oplus V^{j-\frac{1}{2},-j-\frac{1}{2}}.$$
The dimension of such a representation is again $4j+1$.
\item Atypical not-completely reducible - $j=\pm b$:
There are also reducible representations of dimension $8j$ for $j=\pm b$, consisting of four
multiplets of $\su(2)\oplus\u(1)$, which contain the atypical irreducible representation as an
invariant subspace.
$$R^{j,b}=V^{j,b}\oplus V^{j-\frac{1}{2},b-\frac{1}{2}}\oplus V^{j-\frac{1}{2},b+\frac{1}{2}}\oplus V^{j-1,b}.$$
The complementary subset is not itself an invariant subspace, so the representation is not fully reducible.
\end{itemize}
The Casimirs $C_2$ and $C_3$ do not classify the atypical representations.   Evaluating these
operators on the representations, we arrive at $C_2= j^2-b^2$ and $C_3 = b(j^2-b^2)$, meaning that
both are zero on all such representations.  The action of the algebra is given in Appendix
\ref{action}.

Our first thought might be to simply exclude the troublesome atypical representations from the
beginning. But we must be more careful than that since these representations can occur in the
decomposition of the tensor product of two typical representations
\be
R^{j_1,b_1}\otimes R^{j_2,b_2}  = \bigoplus_{j=|j_1-j_2|}^{j_1+j_2}
R^{j,b_1+b_2}\oplus\bigoplus_{j=|j_1-j_2+\frac{1}{2}|}^{j_1+j_2-\frac{1}{2}}
R^{j,b_1+b_2+\frac{1}{2}}\oplus\bigoplus_{j=|j_1-j_2+\frac{1}{2}|}^{j_1+j_2-\frac{1}{2}}
R^{j,b_1+b_2-\frac{1}{2}}\oplus\bigoplus_{j=|j_1-j_2-1|}^{j_1+j_2+1} R^{j,b_1+b_2}
\ee
where the sums over $j$ are in {\it integer} steps.  If  $b_1=j_1-1$ and $b_2 = j_2-1$, we see that
an atypical representations occurs in the first sum on the right hand side.

Therefore, we must place an extra condition on the class of representations allowed to label the
edges of our simplicial complex.  It turns out  that the subset of representations satisfying $\pm
b > j$ forms a subcategory of representations closed under tensor product.   We can also motivate
this condition by noticing that it is exactly these representations upon which we can define a star
operator (with respect to a positive definite scalar product)\cite{Scheunert:1976wj}.  In fact, a
grade star operator does not exist on the representations of $\osp_E(2|2)$ (apart from two special
cases).   Therefore, the parity considerations, which we had to deal with in the $N=1$ case, are no
longer present.\footnote{This means that the superdimension of an $\osp_E(2|2)$ representation is
the same as the dimension.} In fact, the conditions for the existence of a star operator are
stronger still.  It requires $b\in\R$, and that the scalar product reduce to the usual one when
restricted to $\su(2)$.   The effect of the adjoint operation on the algebra is
\be
\begin{array}{lllll}
\textrm{For all $\pm b>j$,} & J_i^\dag = J_i, & B^\dag = B, & &\\
\textrm{for $b> j$,} & Q_+^\dag = \t{Q}_-,  & Q_-^\dag = -\t{Q}_+, &\t{Q}_+^\dag =  Q_-, & \t{Q}_-^\dag = -Q_+,\\
\textrm{for $b< -j$,} & Q_+^\dag = -\t{Q}_-,  & Q_-^\dag = \t{Q}_+, &\t{Q}_+^\dag = -Q_-, & \t{Q}_-^\dag =  Q_+,
\end{array}
\ee
Interestingly, the Bogomol'nyi bound for this algebra is $\{Q_A, \t{Q}_B\}\{Q^A,\t{Q}^B\} =
2(BB-J^iJ_i)\geq 0$.  This imposes the condition $\pm b \geq \sqrt{j(j+1)}$.  This is in fact a
slightly stronger condition than what we have already imposed, and means that the BPS states (which
saturate this bound) lie within the restricted subcategory of representations.

Working with this subcategory of representations closed under tensor product allows to define a 3d
topological state sum based on $\osp_E(2|2)$ following the framework introduced in
\cite{Livine:2003hn,topoinv}. This is our proposal for the spinfoam quantization of $N=2$
3d supergravity.

Moreover, this analysis can be applied to $\osp_E(2|2)$ spin network states for $N=2$ loop quantum
supergravity in the usual $3+1$ spacetime dimensions
\cite{Ling:2000su}. In this context, we can define a notion of BPS states saturating the proposed $|b|\ge j$
bound. It would be interesting to investigate further what consequences this has on the quantum
states describing a supersymmetric black hole in loop gravity.

\begin{figure}[h]
\begin{minipage}{0.13\linewidth}
\psfrag{A}{$R^{j_1,b_1}$}
\psfrag{B}{$R^{j_2,b_2}$}
\psfrag{C}{$R^{j_3,b_3}$}
\includegraphics[width = 0.8\linewidth]{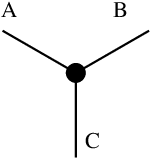}
\end{minipage}\hspace{1cm}
$\left\{\hspace{0.5cm}
\begin{minipage}{0.5\linewidth}
\centering
\psfrag{D}{{\tiny +}}
\psfrag{E}{{\small -}}
\psfrag{F}{{\tiny $\pm$}}
\psfrag{H}{{\tiny $\mp$}}
\psfrag{G}{{\small Pure Gravity}}
\psfrag{arus}{{\small Positively Charged Fermion}}
\psfrag{minus}{{\small Negatively Charged Fermion}}
\includegraphics[width = 0.85\linewidth]{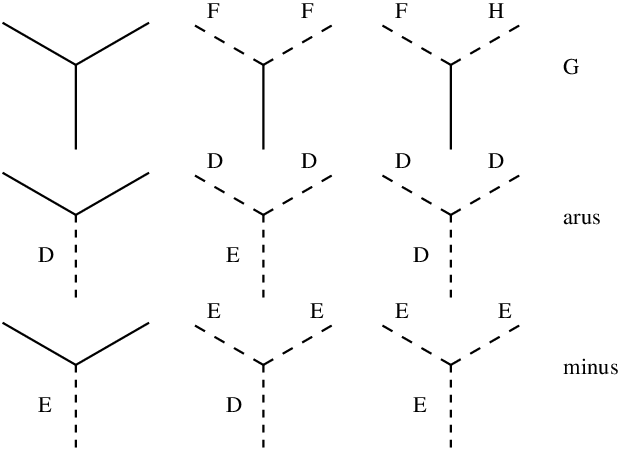}
\end{minipage}
\right.$
\caption{\label{susytwothreej}
The $\osp_E(2|2)$ intertwiner.  The $\su(2)\oplus\u(1)$ representations $V^{j,b}$ and $V^{j-1,b}$
are denoted by solid lines, the $V^{j-\frac{1}{2},b-\frac{1}{2}}$ by $+$-dashed lines, and the
$V^{j-\frac{1}{2},b+\frac{1}{2}}$ by the $-$-dashed lines.  We do not draw all the diagrams;
permutations of the above are possible. }
\end{figure}

To construct the amplitude, we need to label each edge with an orientation.  Then for the each face
of every tetrahedron, we assign an intertwiner $I_f:R^{j_1,b_1}\otimes R^{j_2,b_2}\otimes
R^{j_3,b_3}\rightarrow \C$ or $I_f:R^{j_1,b_1}\otimes R^{j_2,b_2}\otimes (R^{j_3,b_3})^*\rightarrow
\C$ depending on the orientation of the edges.  The intertwiner is unique  up to normalization and
we normalize as before.  To construct the tetrahedral amplitude we contract four such intertwiners.

We illustrate the intertwiner $I_f: R^{j_1,b_1}\otimes R^{j_2,b_2}\otimes R^{j_3,b_3}\rightarrow
\C$  in Figure \ref{susytwothreej}.  Once again, the decomposition of the representations allows
one to distinguish subclasses within the $\osp_E(2|2)$ intertwiner.  Our rationale when classifying
these diagrams is that those marked as pure gravity satisfy $j_1+j_2+j_3 \in \mathbb{N}$.  The
other classes satisfy  $j_1+j_2+j_3 \in \mathbb{N} + \frac{1}{2}$ and
$b_1+b_2+b_3\in\mathbb{Z}+\frac{1}{2}$. In particular, the positively charged fermionic intertwiner
has $b_1+b_2+b_3>0$, while negatively charged one has $b_1+b_2+b_3<0$.  The fermions propagate
along edges of the spin foam $\dman^*_2$.

\begin{figure}[h]
\psfrag{A}{\tiny $F^+$}
\psfrag{B}{\tiny $G$}
\psfrag{C}{\tiny $F^-$}
\psfrag{D}{\tiny $G$}
\psfrag{=}{\footnotesize $=$}
\psfrag{+}{\footnotesize $+$}
\psfrag{do}{\footnotesize$\cdots$}
\psfrag{a}{\tiny $+$}
\psfrag{b}{\tiny $-$}
\psfrag{c}{\tiny $\pm$}
\includegraphics[width = 16cm]{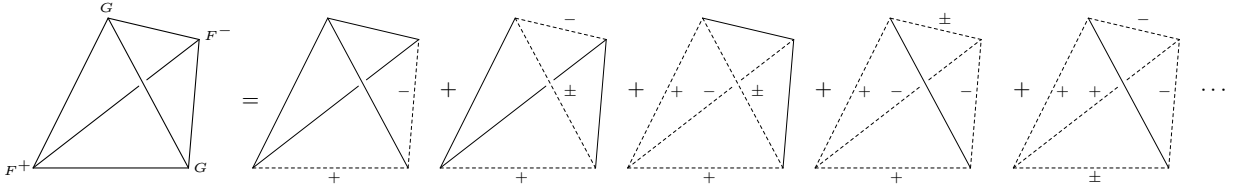}
\caption{\label{sixj2} The $(G,G,F^+,F^-)$ class of diagrams.  $G$  stands for the bosonic intertwiners, while $F^\pm$ represents the $\pm$-fermionic intertwiners.  There are five diagrams occurring in this decomposition up to permutations.}
\end{figure}

The tetrahedral amplitude is a $\{6j,6b\}_{\osp(2|2)}$-symbol meaning that each term in the sum is
the product of a $\{6j\}_{\su(2)}$ and a $\{6b\}_{\u(1)}$.      We separate its constituent
diagrams into nine classes:  $(G,G,G,G)$, $(G,G,F^+,F^+)$, $(G,G,F^+,F^-)$, $(G,G,F^-,F^-)$,
$(F^+,F^+,F^+,F^+)$, $(F^+,F^+,F^+,F^-)$, $(F^+,F^+,F^-,F^-)$, $(F^+,F^-,F^-,F^-)$,
$(F^-,F^-,F^-,F^-)$, where $G$ stands for a bosonic intertwiner, $F^{\pm}$ stands for a
$\pm$-charged fermionic intertwiner. We illustrate one class of diagrams in Figure \ref{sixj2}. It
is interesting to note that there are such terms as $(G,G,F^+,F^-)$.  As we sum over both
orientations, this language represents two processes.  The first is as the annihilation of an
oppositely charged fermion pair.  The second is as the interaction of a positively charged fermion
with the dynamical \lq\lq background"  $\su(2)\oplus\u(1)$ gauge field to produce a negatively
charged fermion.

Finally, the amplitude (\ref{part2})  represents the propagation of charged particles in a
dynamical geometry.  The fermion paths form a Feynman graph embedded into the spin foam.
Topologically, the Feynman graphs are the same as in the $N=1$ case, but here they contain more
information.  The particles are charged under an extra $\u(1)$ algebra.  Essentially the particles
only propagate, there only interaction is with the gauge field in order .  This mimics the
classical action (which is cubic and so could not contain the interaction of spin-$\frac{1}{2}$
fermions) where one notices the $gauge-fermion-antifermion$ interaction terms.

\subsection{The Lorentzian Case}

Up to this point we have only described the case of $3d$ Riemannian gravity, but a more physically
interesting case is 2+1-$d$ gravity where the underlying symmetry algebra is $\su(1,1)$.   In the
supersymmetric context, we examine the case of a spin foam model based on $\osp_L(2|2)$ where the
bosonic subalgebra is $\su(1,1)\oplus\u(1)$.\footnote{The $N=1$ case was dealt with in
\cite{Livine:2003hn}.}  We arrive at the generators for this algebra through rotating by $i$ the
generators $J_\pm$ of its Riemannian counterpart $\osp_E(2|2)$.\footnote{The Hermiticity relations
for $\su(1,1)$ are $J_{\pm}^\dag = -J_\mp$ and $J_3^\dag = J_3$.} The only changes in the
commutation relations (\ref{susy2cc}) are
\be
\begin{array}{lll}
[J_{\pm},Q_{\mp}]=-iQ_{\pm},&[J_{\pm},\tilde{Q}_{\mp}]=-i\tilde{Q}_{\pm},& \{Q_{\pm},\tilde{Q}_{\pm}\}={\pm}iJ_{\pm}.
\end{array}
\ee
The representations of $\osp_L(2|2)$ can once again be decomposed into a direct sum of
representations of its bosonic subalgebra. We restrict to the unitary principal representations of
$\su(1,1)$  which are of two kinds: a continuous series labeled by $s\in\mathbb{R}$ and two
discrete series labeled by $j\in\frac{1}{2}\mathbb{N}$, one positive and one negative.  The basis
for the continuous series is the set of vectors $|s, j_3>$ where either $j_3\in\mathbb{Z}$ or
$j_3\in\mathbb{Z}+\frac{1}{2}$.  The basis for the positive discrete series is the set of vectors
$|j,j_3>$ with $j_3\geq j$ and $j_3\in\frac{1}{2}\mathbb{N}$, while for the negative series it is
the set of vectors $|j,j_3>$ with $j_3\leq -j$ and $-j_3\in\frac{1}{2}\mathbb{N}$.   The Casimir
evaluated on these representations (in its capacity kinematical length-squared operator) suggests
that the continuous representations should label space-like edges while the positive and negative
discrete series should label the future time-like and past time-like edges respectively
\cite{3dlqg}.\footnote{The Casimir for $\su(1,1)$ takes the values:  $s^2+\frac{1}{4}$ for the continuous
series and  $-j(j-1)$ for the discrete series.  They are of opposite sign.}    In $2+1$-d quantum
gravity, we have a choice when proposing a topological state sum:  we can choose to sum over all
the principal representations, or we can restrict the sum to just the positive discrete series.
Interestingly, the set of positive discrete representations is a subcategory closed under tensor
product.  A simplicial complex labeled in either way will be a topological invariant.  We will
restrict to this class of representations for the $\su(1,1)$ subalgebra of $\osp_L(2|2)$ also. With
this proviso, the representation theory of $\osp_L(2|2)$ has a very similar appearance to that of
$\osp_E(2|2)$. The representations are $R^{j,b}$ where $j$ is the isospin and $b\in\C$ is the
charge.   They once again fall into four categories, one typical and three atypical. Under tensor
product however, the representations satisfy
\be
R^{j_1,b_1}\otimes R^{j_2,b_2}  = \bigoplus_{j>j_1+j_2}
R^{j,b_1+b_2}\oplus\bigoplus_{j>j_1+j_2-\frac{1}{2}}
R^{j,b_1+b_2+\frac{1}{2}}\oplus\bigoplus_{j>j_1+j_2-\frac{1}{2}}
R^{j,b_1+b_2-\frac{1}{2}}\oplus\bigoplus_{j>j_1+j_2+1} R^{j,b_1+b_2}
\ee
This time around, the restriction to a consistent subcategory of representations is $|b| < j$,
which is simply the reverse inequality compared to the Riemannian theory.


Now that we have successfully defined a consistent representation theory, we can proceed in an
analogous fashion to the Riemannian scenario and define a spin foam model based on the restricted
subcategory of representations.  The state sum will be triangulation independent and by decomposing
the representations into their isospins, we can interpret certain terms in the sum as representing
charged fermions along edges of the spin foam.

\subsection{An alternative state sum construction}

The fact that a $\osp_E(1|2)$ irreducible representation is made of two $\su(2)$ spin
representations suggests another strategy to derive spinfoam models with higher supersymmetry. One
could try to pile $\su(2)$ representations together in order to form supersymmetric multiplets. For
example, we can consider the reducible representations made as the direct sum of three copies of
$\su(2)$,
\be
R^{j} = V^{j}\oplus V^{j-\frac{1}{2}}\oplus V^{j-1}
\ee
Indeed, this satisfies the relation
\be
R^{j}\otimes R^k = \bigoplus_{l = |j-k|}^{j+k}R^l \oplus \bigoplus_{l =
|j-k+\frac{1}{2}|}^{j+k-\frac{1}{2}} R^l\oplus\bigoplus_{l=|j-k+1|}^{j+k-1} R^l
\ee
under tensor product.  This category of representations is closed under tensor product, so we can
build a topological state sum based on them. This actually generalizes to stacks of $\su(2)$
representations of arbitrary size.

On the other hand, the task now is to identify an algebra to which the $R^{(j)}$'s provide faithful
(irreducible) representations. Since they are made of three $\su(2)$ representations, it could be
interpreted as the representation of some algebra in between $N=1$ and $N=2$ supergravity theories.
Actually, it seems to correspond to the $\osp_E(2|2)$ algebra where we would have gauge-fixed the
$\u(1)$ generator $B$ to $b=0$. Then the quadratic Casimir of the resulting algebra is $C_2 =
\eta^{ij}J_iJ_j + Q_A\epsilon^{AB}\t{Q}_B$ which yields $C_2 = j^2$ on a representation. This
characterizes the representation uniquely.

The next difficulty is to find the classical theory from which one could derive this model. Our
guess is simply
\be
\cS_{alt}[\cE,\cA] = \int_\cM\Str(\cE\wedge\cF[\cA]),
\ee
where we have removed the $\u(1)$ component of the superfields,
\be
\cE  =  E^iJ_i + \phi^AQ_A + \t{\phi}^A\t{Q}_A, \hspace{1cm}   \cA  =  W^iJ_i + \psi^AQ_A + \t{\psi}^A\t{Q}_A,
\ee
Written out explicitly in components of the multiplets $\cE$ and $\cA$, this action can be
rewritten as
\be
\cS_{alt}[\cE, \cA] = \int_{\cM} \left\{ \Str(E\w (F[W] + \psi\wedge\t{\psi}) + \t{\phi}\w D\psi +\phi \w D\t{\psi} \right\}
\ee
where $D = \pd  + W\w$ is the covariant derivative with respect to gravity and $F(W) = dW + W\w W$
is the gravitational curvature. This spin foam model would then give configurations for gravity
plus two indistinguishable fermion types.

\section{Conclusion}

In this paper, we reviewed the relevant aspects of $N=1$ quantum supergravity in the three
dimensions and extended the theory to the $N=2$ case.  This is a richer theory as it contains a
$\u(1)$ gauge theory.   We constructed a spin foam based on the gauge algebra $\osp(2|2)$.  As with
$N=1$ supergravity, the classical action contains fermionic degrees of freedom.  We must identify
these properties in the quantum theory.  The super Ponzano-Regge state sum contains configurations
that do not arise in the $\su(2)$ quantum gravity state sum.   More precisely, the difference is
that in the supergravity theories, the triple of isospins labeling the three edges of a triangle
need not necessarily satisfy $j_1+j_2+j_3\in \N$.  The intertwiner of such a triple is viewed as a
spin-$\frac{1}{2}$ fermion propagating along an edge of the spin foam.  When the gauge algebra is
$\osp(1|2)$ there is one type of fermion, while for the $\osp(2|2)$, we have positively and
negatively charged fermions.

This interpretation is not steadfast, however, and we need to solidify our reasoning with an
analysis of the pertinent semi-classical limit.  This will be the subject of later work
\cite{Livine}.  We shall develop a connection between the model developed here and fermionic fields
coupled to gravity in the arena of Regge calculus. The strategy will be to analyze the asymptotic
behavior of the $N=1$ and $N=2$ supersymmetric $\{6j\}$ symbols. Indeed, we already know how the
standard $\{6j\}$ symbol for $\su(2)$ is related to the Regge action for 3d gravity. Factorizing
out this term in the supersymmetric $\{6j\}$ symbols, we hope to identify the discrete path
integral amplitude describing the dynamics of the supersymmetric fields coupled to gravity. Within
the spinfoam graviton propagator framework \cite{graviton}, this will allow to compute the
scattering amplitudes for fermions and $\u(1)$ gauge fields (for $N=2$) coupled to 3d quantum
gravity.

One should also study whether it is possible to extend the state sum in the Lorentzian case to
include the continuous representations, and to check the conditions required to allow the existence
of a star operator.  The continuous representations label space-like edges and so are necessary to
define a spatial hypersurface.

The extension to higher BF-theory could be seen as the first link in the chain connecting this work
to supergravity, which occurs as a constrained $BF$ theory \cite{Ling:2000ss}, and with a possible
non-perturbative background independent definition of M-theory following the logic of
\cite{Mtheory}.

\vspace{1cm}

\noindent {\it Acknowledgements:}
Research at Perimeter Institute for Theoretical Physics is supported
in part by the Government of Canada through NSERC and by the Province of Ontario through MRI.


\begin{appendix}
 \renewcommand{\theequation}{\thesection.\arabic{equation}}

\section{Action on a typical representation}
\label{action}
The action of the operators on the representations is (for
$R^{j,b}$, $j\neq\pm b$):
\be
J_{\pm}|j,j_3,b>=\sqrt{(j\mp j_3)(j\pm j_3+1)}|j,j_3\pm 1,b>,\qquad
J_3|j,j_3,b>=j_3|j,j_3,b>,\qquad
B|j,j_3,b>= b|j,j_3,b>,\nonumber
\ee
\begin{eqnarray*}
Q_{\pm}|j,j_3,b>&=&\pm\sqrt{\frac{(b+j)(j\mp
j_3)}{2j}}|j-\frac{1}{2},j_3\pm\frac{1}{2},b+\frac{1}{2}>\\
\tilde{Q}_{\pm}|j,j_3,b>&=&\pm\sqrt{\frac{(b-j)(j\mp
j_3)}{2j}}|j-\frac{1}{2},j_3\pm\frac{1}{2},b-\frac{1}{2}>\\
&&\\ &&\\
Q_{\pm}|j-\frac{1}{2},j_3,b-\frac{1}{2}>&=&\sqrt{\frac{(b-j)(j\pm
j_3+\frac{1}{2})}{2j}}|j,j_3\pm\frac{1}{2},b>
\mp\sqrt{\frac{(b+j)(j\mp
j_3-\frac{1}{2})}{2j}}|j-1,j_3\pm\frac{1}{2},b>\\
\tilde{Q}_{\pm}|j-\frac{1}{2},j_3,b-\frac{1}{2}>&=&0\\
&&\\ &&\\
Q_{\pm}|j-\frac{1}{2},j_3,b+\frac{1}{2}>&=&0\\
\tilde{Q}_{\pm}|j-\frac{1}{2},j_3,b+\frac{1}{2}>&=&\sqrt{\frac{(b+j)(j\pm
j_3+\frac{1}{2})}{2j}}|j,j_3\pm\frac{1}{2},b>
\pm\sqrt{\frac{(b-j)(j\mp
j_3-\frac{1}{2})}{2j}}|j-1,j_3\pm\frac{1}{2},b>\\
&&\\ &&\\
Q_{\pm}|j-1,j_3,b>&=&\sqrt{\frac{(b-j)(j\pm
j_3)}{2j}}|j-\frac{1}{2},j_3\pm\frac{1}{2},b+\frac{1}{2}>\\
\tilde{Q}_{\pm}|j-1,j_3,b>&=&-\sqrt{\frac{(b+j)(j\pm
j_3)}{2j}}|j-\frac{1}{2},j_3\pm\frac{1}{2},b-\frac{1}{2}>\\
\end{eqnarray*}
The value assumed by the Casimirs acting on each state of the representation $R^{j,b}$ are:
\be C_2:\quad j^2-b^2 \quad\quad C_3: \quad b(j^2-b^2).\ee
It is clear that they both vanish for $j=\pm b$, so they do not pick out these representations.  The states are normalised.

\end{appendix}

\end{document}